\documentclass{ctr_summer}

\usepackage{ctrfont}

\usepackage{epsfig}
\usepackage{psfrag}
\usepackage{amsmath,rotating}
\usepackage{color}

\allowdisplaybreaks

\renewcommand{\thesection}{\arabic{section}}

\newcounter{eee}

\newcounter{eeeb}

\renewcommand{\x}{\cdot}

\newcommand{\dd}{\partial}
\newcommand{\de}{{\rm \, d}}

\renewcommand{\vec}[1]{\mbox{\boldmath $ #1$}}

\newcommand{\R}{\mathrm{R}}
\renewcommand{\P}{\mathrm{P}}

\newcommand{\Pm}{\mathrm{P}_\mathrm{m}}

\title{Turbulent three-dimensional MHD dynamo model in spherical
  shells: Regular oscillations of the dipolar field} 
\shorttitle{Dipolar dynamo oscillations}

\author{ R.~D.~Simitev$^\ast$\footnotetext{$^\ast$ School of
  Mathematics and Statistics, University of Glasgow, Glasgow, UK}, 
  F.~H.~Busse$^\S$\footnotetext{$^\S$ Institute of Physics, University of Bayreuth,
  Bayreuth, Germany}  \and
  A.~G.~Kosovichev$^\ddag$\footnotetext{$^\ddag$Hansen Experimental Physics
  Laboratories, Stanford University, USA},
  }

\shortauthor{Simitev \textit{et al.}}

\begin{document}

\setcounter{page}{475}

\maketitle

We report the results of three-dimensional numerical simulations of
convection-driven dynamos in relatively thin rotating spherical shells
that show a transition from an strong non-oscillatory dipolar magnetic
field to a weaker regularly oscillating dipolar field. The transition
is induced primarily by the effects a stress-free boundary
condition. The variation of the inner to outer radius ratio is found
to have a less important effect. 
 
\vskip0.1in
\hrule

\section{Introduction}
\label{s:intro}

The Sun possesses a deep layer of intensely turbulent convection near
its surface.  The dynamical properties of this 
zone continue to elude basic physical explanation. The most important
questions involve the solar differential rotation profile and the
periodicity of the solar magnetic activity. The differential rotation
is known from helioseismological studies (e.g.~\cite{Schou}):  
the angular velocity variation observed at the surface where the
rotation is faster near the equator and slower near the poles, extends
through the convection zone with little radial dependence.  
The basic feature of the solar activity is the existence of a 
periodic 22-year long cycle which is mainly manifested in the periodic
appearance of sunspots accompanied by an oscillation of the
antisymmetric (dipolar) parts of the solar magnetic field during which it
reverses polarity (e.g.~\cite{Brandenburg}).
Both the differential rotation and the solar activity cycle are
persistent global scale features of the Sun and it is difficult
to understand how they emerge from the intensely turbulent flow in the
solar convection zone which includes scales ranging from granules (1
Mm in horizontal size), to supergranules (30 Mm), and possibly giant
cells (over 200 Mm) and which shows similar contrasts in time scales. 
A variety of modeling approaches ranging from linear theory
(e.g.~\cite{Rudiger}) and weakly nonlinear theory (\cite{Busse1970}) to high-resolution
three-dimensional global simulations of compressible convective dynamos
in spherical shells (\cite{Brun}, \cite{Browning})
have been employed but the precise mechanisms leading to the peculiar
form of the profile of solar differential rotation, and the
regularity of its magnetic activity remain unclear.  

In a recent study \cite{Dormy} carried out
three-dimensional numerical simulations of self-excited convective
dynamos  with the aim of exploring the dependence of solutions on the
thickness of the spherical shell. They employed a model that has been  
used as a benchmark by the geodynamo community (\cite{Christensen}), 
which they modified by replacing the no-slip boundary conditions
at the top of the spherical shell by stress-free ones. They 
found that a transition occurs from non-oscillating to regularly
oscillating dipolar dynamo solutions as the thickness of the spherical
shell is decreased and concluded that this is a purely geometric
effect. They discussed their finding in the light of the facts that
the outer core, the layer where the non-oscillatory geodynamo is
believed to be generated is a relatively thick layer in the case of the Earth,
whereas the solar convective zone is a relatively thin.

However, it is known that oscillating dipolar magnetic
solutions can also be obtained in dynamos in thick spherical shells
(Busse \& Simitev 2006), and it remains unclear whether the transition
found by Goudard \& Dormy (2008) is the result of variation of the shell
thickness or the introduction of a stress-free boundary condition.
In this report we attempt to clarify this question and extend the
results of Goudard \& Dormy (2008) by exploring the effects of a
different set of boundary conditions. We find that the transition to 
an oscillatory state is more likely due to the use of stress-free 
boundaries.

\section{Mathematical formulation of the problem and methods of solution}

We consider a spherical fluid shell rotating about a fixed vertical
axis. We assume that a  
static state exists with the temperature distribution
$$ 
%T_S = T_0 - \beta d^2 r^2 /2 + \Delta T \eta r^{-1} (1-\eta)^{-2},
T_S = T_0 + \Delta T \eta r^{-1} (1-\eta)^{-2},
$$
where $r$ denotes the distance from the center of the spherical shell,
$\eta$ denotes the ratio of inner to outer radius of the shell and $d$
is its 
thickness, and $\Delta T$ is the temperature difference between the
boundaries.
%$T_S = T_0 - \beta d^2 r^2 /2$. Here $rd$ is the length of
%the position vector with respect to the center of the sphere.
The gravity field is given by 
$$
\vec g = - d \gamma \vec r.
$$
In addition to  $d$, the
time $d^2 / \nu$,  the temperature $\nu^2 / \gamma \alpha d^4$, and 
the magnetic flux density $\nu ( \mu \varrho )^{1/2} /d$ are used as
scales for the dimensionless description of the problem  where $\nu$ denotes
the kinematic viscosity of the fluid, $\kappa$ its thermal diffusivity,
$\varrho$ its density and $\mu$ is its magnetic permeability.
The equations of motion for the velocity vector $\vec u$, the heat
equation for the deviation 
$\Theta$ from the static temperature distribution, and the equation of
induction for the magnetic flux density $\vec B$ are thus given by 
\begin{subequations}
\begin{gather}
\label{1a}
\partial_t \vec{u} + \vec u \cdot \nabla \vec u + \tau \vec k \times
\vec u = - \nabla \pi +\Theta \vec r + \nabla^2 \vec u + \vec B \cdot
\nabla \vec B, \\
\label{1b}
\nabla \cdot \vec u = 0, \\
\label{1c}
\P(\partial_t \Theta + \vec u \cdot \nabla \Theta) = (\R\, \eta
r^{-3} (1 - \eta)^{-2}) \vec r \cdot \vec u + \nabla^2 \Theta, \\
\nabla \cdot \vec B = 0, \\
\label{1d}
\nabla^2 \vec B =  \Pm(\partial_t \vec B + \vec u \cdot \nabla \vec B
-  \vec B \cdot \nabla \vec u),
\end{gather}
\end{subequations}
where $\partial_t$ denotes the partial derivative with respect to time
$t$ and where all terms in the equation of motion that can be written
as gradients have been combined into $ \nabla \pi$. The Boussinesq
approximation has been assumed in that the density $\varrho$ is
regarded as constant except in the gravity term where its temperature
dependence, given by $\alpha \equiv - ( \de \varrho/\de T)/\varrho =${\sl
const}, is taken into account. The Rayleigh numbers $\R$,
the Coriolis number $\tau$, the Prandtl number $\P$ and the magnetic
Prandtl number $\Pm$ are defined by 
%3
\begin{equation}
%\R = \frac{\alpha \gamma \beta d^6}{\nu \kappa} , 
\R = \frac{\alpha \gamma \Delta T d^4}{\nu \kappa},
\enspace \tau = \frac{2
\Omega d^2}{\nu} , \enspace \P = \frac{\nu}{\kappa} , \enspace \Pm = \frac{\nu}{\lambda},
\end{equation}
where $\lambda$ is the magnetic diffusivity.  Because the velocity 
field $\vec u$ as well as the magnetic flux density $\vec B$ are
solenoidal vector fields,   the general representation in terms of
poloidal and toroidal components can be used 
\begin{subequations}
\begin{gather}
\vec u = \nabla \times ( \nabla v \times \vec r) + \nabla w \times 
\vec r \enspace , \\
\vec B = \nabla \times  ( \nabla h \times \vec r) + \nabla g \times 
\vec r \enspace .
\end{gather}
\end{subequations}
By multiplying the (curl)$^2$ and the curl of equation \eqref{1a} by
$\vec r$ we obtain two equations for $v$ and $w$  
%2a
\begin{subequations}
\label{momentum}
\begin{gather}
[( \nabla^2 - \partial_t) {\cal L}_2 + \tau \partial_{\varphi} ] \nabla^2 v +
\tau {\cal Q} w - {\cal L}_2 \Theta  
= - \vec r \cdot \nabla \times [ \nabla \times ( \vec u \cdot
\nabla \vec u - \vec B \cdot \nabla \vec B)], \\
[( \nabla^2 - \partial_t) {\cal L}_2 + \tau \partial_{\varphi} ] w - \tau {\cal Q}v 
= \vec
r \cdot \nabla \times ( \vec u \cdot \nabla \vec u - \vec B \cdot
\nabla \vec B), 
\end{gather}
\end{subequations}
where $\partial_{\varphi}$ denotes the partial derivative with respect to
the angle $\varphi$ of a spherical system of coordinates $r, \theta, \varphi$
and where the operators ${\cal L}_2$ and $\cal Q$ are defined by 
\begin{gather}
{\cal L}_2 \equiv - r^2 \nabla^2 + \partial_r ( r^2 \partial_r), \nonumber\\
{\cal Q} \equiv r \cos \theta \nabla^2 - ({\cal L}_2 + r \partial_r ) ( \cos \theta
\partial_r - r^{-1} \sin \theta \partial_{\theta}). \nonumber
\end{gather}
The heat equation for the dimensionless deviation $\Theta$ from the
static temperature distribution can be written in the form
%2c
\begin{equation}
\label{heat}
\nabla^2 \Theta + (\R\, \eta
r^{-3} (1 - \eta)^{-2}) {\cal L}_2 v = \P ( \partial_t + \vec u \cdot \nabla ) \Theta,
\end{equation}
and the equations for $h$ and $g$ are obtained through the multiplication of
equation \eqref{1d} and of its curl by $\vec r$
\begin{subequations}
\label{induction}
\begin{gather}
\nabla^2 {\cal L}_2 h = \Pm [ \partial_t {\cal L}_2 h - \vec r \cdot
\nabla \times ( \vec u \times \vec B )], \\
\nabla^2 {\cal L}_2 g = \Pm [ \partial_t {\cal L}_2 g - \vec r \cdot
\nabla \times ( \nabla \times ( \vec u \times \vec B ))].
\end{gather}
\end{subequations}

In this report we discuss results obtained with two different
sets of velocity boundary conditions, namely no-slip conditions
given by
\begin{gather}
\label{ns}
v = \partial_r v = w = 0
\qquad \mbox{ at } r=r_i \equiv \eta/(1-\eta) \mbox{ and } r=r_o
\equiv 1/(1-\eta),
\end{gather}
and mixed conditions, i.e.~a combination of no-slip conditions at
the inner spherical boundary and a stress-free condition at the outer
boundary, given by 
\begin{gather}
\label{mixed}
v = \partial_r v = w = 0
\qquad \mbox{ at } r=r_i \equiv \eta/(1-\eta),  \\
v = \partial^2_{rr}v = \partial_r (w/r) = 0
\qquad \mbox{ at } r=r_o \equiv 1/(1-\eta).\nonumber
\end{gather}
As boundary conditions for the heat equation, we assume fixed temperatures
\begin{equation}
\label{hbc}
\hspace*{-8mm}
\Theta = 0 
\qquad \mbox{ at } r=r_i \equiv \eta/(1-\eta)  \mbox{ and } r=r_o \equiv 1/(1-\eta).
\end{equation}
For the magnetic field, electrically insulating
boundaries are assumed such that the poloidal function $h$ must be 
matched to the function $h^{(e)}$, which describes the  
potential fields 
outside the fluid shell  
%4b
\begin{equation}
\hspace*{-8mm}
\label{mbc}
g = h-h^{(e)} = \partial_r ( h-h^{(e)})=0 
\mbox{ at } r=r_i \equiv \eta/(1-\eta)  \mbox{ and } r=r_o \equiv 1/(1-\eta).
\end{equation}
But computations for the case of an inner boundary with no-slip
conditions and an electrical conductivity equal to that of the fluid
have also been done. The numerical integration of equations
\eqref{momentum},\eqref{heat}, and \eqref{induction} together with boundary
conditions \eqref{hbc}, \eqref{mbc} and any of \eqref{ns} or
\eqref{mixed}, proceeds with the pseudo-spectral 
method as described by Tilgner (1999)
which is based on an expansion of all dependent variables in
spherical harmonics for the $\theta , \varphi$-dependences, i.e. 
%5
\begin{equation}
v = \sum \limits_{l,m} V_l^m (r,t) P_l^m ( \cos \theta ) \exp \{ im \varphi \}
\end{equation}
and analogous expressions for the other variables, $w, \Theta, h$ and $g$. 
Here $P_l^m$ denotes the associated Legendre functions.
For the $r$-dependence expansions in Chebychev polynomials are used. 
For the computations to be reported in the following a minimum of
41 collocation points in
the radial direction and spherical harmonics up to the order 128 have been
used.

\section{Dependence of convection and dynamo solutions on the
  thickness of the spherical shell}

In order to clarify whether the transition from non-oscillatory to
oscillatory dynamos reported by Goudard \& Dormy (2008) is the
exclusive result of variation in the radius ratio $\eta$, we started with
a case similar to the one described in their work and identical
to the geodynamo benchmark simulation described in (Christensen \textit{et
al.} 2001). More precisely, we employ no-slip boundary conditions, as given
in the preceding section, both at the inner and outer surface of the 
shell. We then simulate a sequence of dynamo solutions where we
increase the value of $\eta$  while keeping all other parameter values
fixed. 

Global quantities that characterize the basic properties of dynamo
solutions include the magnetic energy density that can be decomposed
into several components, namely poloidal and toroidal energy densities,
each of which can be further decomposed into mean and fluctuating
parts. The corresponding definitions are given by the expressions 
\begin{gather}
\overline{M}_p = \frac{1}{2} \langle \mid\nabla \times ( \nabla
\overline{h}
\times \vec r )\mid^2 \rangle ,  \quad
 \overline{M}_t = \frac{1}{2} \langle \mid\nabla
\overline g \times \vec r \mid^2 \rangle, \nonumber\\
\widetilde{M}_p = \frac{1}{2} \langle \mid\nabla \times ( \nabla
\widetilde h
\times \vec r )\mid^2 \rangle , \quad
 \widetilde{M}_t = \frac{1}{2} \langle \mid\nabla
\widetilde g \times
\vec r \mid^2 \rangle, \nonumber
\end{gather}
where $\langle\cdot\rangle$ indicates the average over the fluid shell
and $\overline h$ refers to the axisymmetric component of $h$.
$\widetilde h$ is defined by $\widetilde h = h - \overline h $.
Similarly, kinetic energy densities $\overline{E}_p$,
$\overline{E}_t$, $\widetilde{E}_p$, and $\widetilde{E}_t$
can be defined with analogous expressions where $v$ and $w$ replace
$h$ and $g$.
In addition, the energy densities can be divided into those of
fields that are antisymmetric (axial dipole symmetry) and those that
are symmetric (axial quadrupole symmetry) with respect to the
equatorial plane. The dipole (quadrupole) fields are described  by spherical
harmonic $Y_l^m$ with odd (even) $l+m$.
\begin{figure}
\vspace*{5mm}
\psfrag{Ex, Mx}{$E_x,M_x$}
\psfrag{eta}{$\eta$}
%Data in 
%          /mnt2/dat/dyn/exxp1t2r100000m1p5ns_engs
%          /mnt2/dat/dyn/exxp1t2r100000m1p5ns_engs/exxp1t2r100000m1p5ns.agr
%          /mnt2/dat/dyn/exxp1t2r100000m1p5ns_engs/bwexxp1t2r100000m1p5ns.agr
\begin{center}
\epsfig{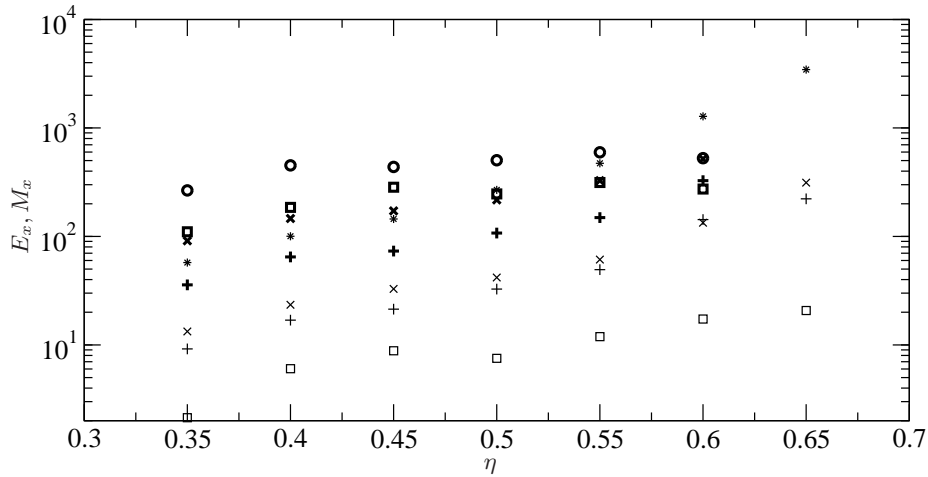}
\end{center}
%\vspace{0mm}
\caption{The averaged magnetic, $M$, and kinetic, $E$,
energy density components, and the helicity, He, as a function of the
radius ratio $\eta$ for fixed values of the other parameters: $\P=1$,
$\R=10^5$, 
$\tau=2000$, $\Pm=5$ and no-slip velocity boundary conditions. 
The components $\overline{X}_p$, $\overline{X}_t$,
$\hat X_p$, and $\hat X_t$ are shown by crosses, squares,
triangles and circles, respectively, where $X=M, E$.  The
energy density $\overline{E}_p$, has not been included because it is
nearly two orders of magnitude smaller than $\overline{E}_t$. The
values of helicity are indicated by stars.
}
\label{fig01}
\end{figure}
\begin{figure}
\vspace*{5mm}
\begin{center}
%Data in 
%          /mnt2/dat/dyn/ ... individual cases
%          /mnt2/dat/dyn/exxp1t2r100000m1p5ns_engs/l__plots__/col
%          /mnt2/dat/dyn/exxp1t2r100000m1p5ns_engs/l__plots__/col/3mmm.tex
\epsfig{file=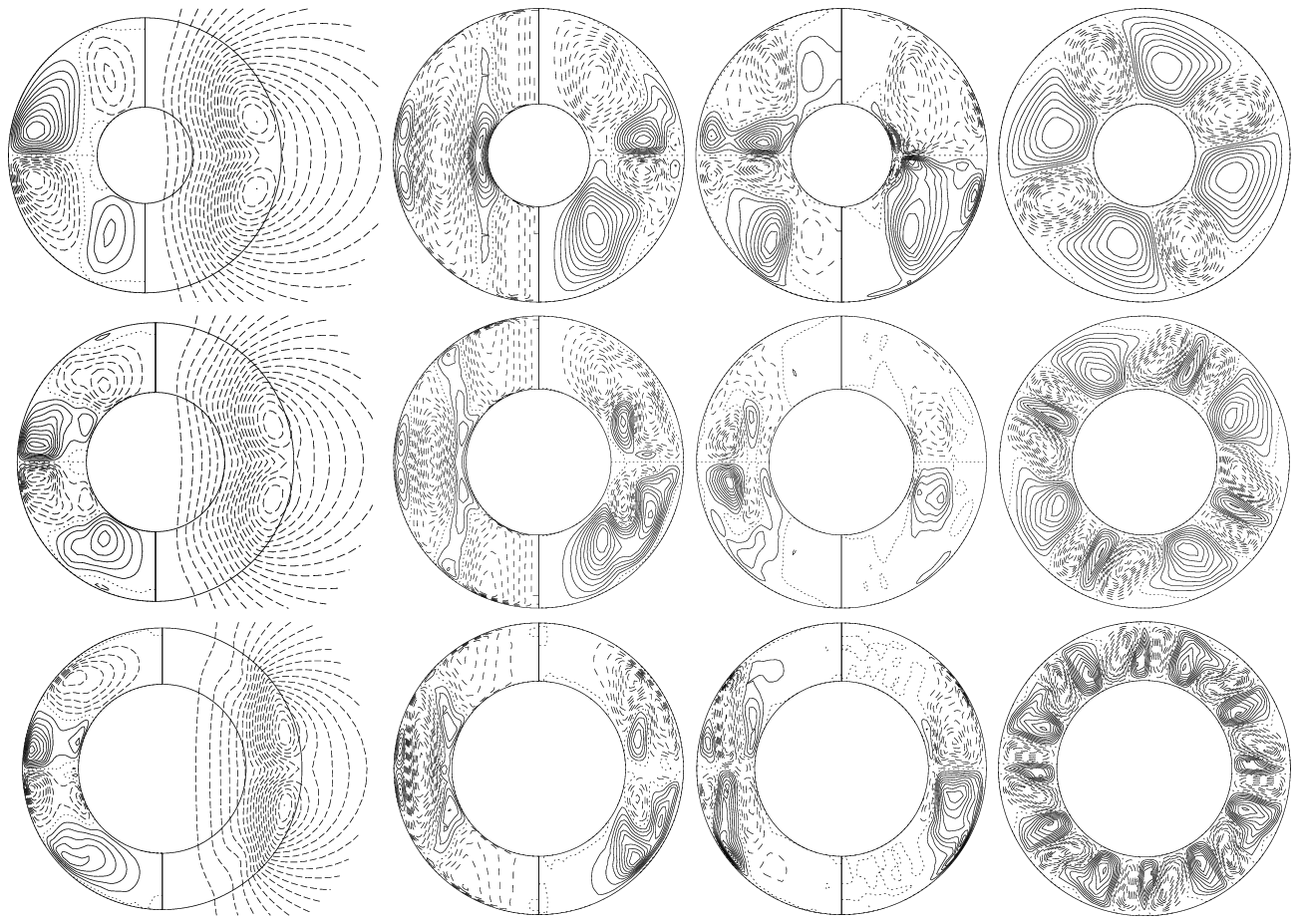,width=\textwidth,clip=}
\end{center}
\vspace{3mm}
\caption{Snapshots of selected components of the dynamo
solutions with $\P=1$, $\R=10^5$, $\tau=2000$, $\Pm=5$, and no-slip
velocity boundary conditions for varying radius ratio $\eta$. 
  Each circle of the first column shows lines of constant
  $\overline{B_{\varphi}}$ in the left half and of $r \sin \theta
  \dd_\theta   \overline{h}=$ const.~in the right half. 
%The second
%  column shows meridional lines of constant $B_r$ at $r=r_o$. 
The
  second column shows meridional lines of constant
  $\overline{u}_\varphi$ in the left half and of $r \sin \theta
  \dd_\theta \overline{v}$ in the right half. The third column shows
  meridional lines of constant values of the azimuthally averaged
  cross-helicity in the left half and helicity in the right
  half. Each plot in the last column shows poloidal streamlines $r
  \partial_\varphi v =$ const.~in the 
  equatorial plane. The value of $\eta=$ 0.35, 0.5, 0.6 in the first,
  second, and third row, respectively.
}
\label{fig02}
\end{figure}
Other global quantities of interest are the helicity defined as
$$
\mathrm{He} = \langle(\nabla \times \vec{u})\cdot\vec{u}\rangle,
$$
and the cross-helicity which is given by 
$$
\mathrm{XHe} = \langle\vec{u}\cdot\vec{B}\rangle.
$$
In Fig.~\ref{fig01} these first-order characteristics of dynamo
solutions are plotted as a function of the radius ratio $\eta$ of the
spherical shell, whereas in Fig.~\ref{fig02} snapshots of typical
spatial solution structures are presented. We remark that a value of
$\eta=0.35$ is typically assumed to be relevant in the case of the
Earth, where the inert inner core extends to less than 40\% of the
core radius; a value of $\eta=0.7$ is appropriate for the Sun,
where the radiative zone fills about 70\% of the solar radius. 
The well-known benchmark solution of (Christensen \textit{et al.}, 2001) has
$\eta=0.35$ and represents a strong dipole. In Fig.~\ref{fig01} this
is evident from the fact that the mean 
poloidal dipolar magnetic energy $\overline{M}_p$ makes the dominant
contribution. This solution is known to be quasi-stationary in
that the energy densities remain constant in time, and the only
time dependence is the drift of the convection pattern in the azimuthal
direction. The solutions at larger values of the radius ratio $\eta$
exhibit a similar behavior, namely they remain strong
non-oscillatory dipoles. All components of the kinetic energy, as well
as the helicity, grow by roughly an order of magnitude as $\eta$
increases from 0.35 to 0.65. The magnetic energy components grow
only slightly despite the increasingly vigorous convective flow, and at
$\eta=0.65$ dynamo action is lost indicating that as the shell
thickness decreases it is more difficult to sustain magnetic field
generation. The relative contributions of the various magnetic energy
components remain unchanged. An inspection of Fig.~\ref{fig02} indicates
 that the main effects of increasing $\eta$ are the growth of the
wavenumber of convection and the migration to lower latitudes of
both, convective motions and magnetic fields, while the polar
regions gradually become less active.  To ensure that the described
behavior is not transient we have continued the simulations for more
than 20 magnetic diffusion times in each case.

\begin{figure}
\vspace*{5mm}
\psfrag{E}{$E$}
\psfrag{Mdip}{$M_\mathrm{dip}$}
\psfrag{HG}{$H_{1,2}^0$}
%Data in 
%          /mnt2/dat/dyn/e065p1t2r100000m1p4.5mvbcB.period/ens
%          /mnt2/dat/dyn/e065p1t2r100000m1p4.5mvbcB.period/ens/Engs.agr
\begin{center}
\epsfig{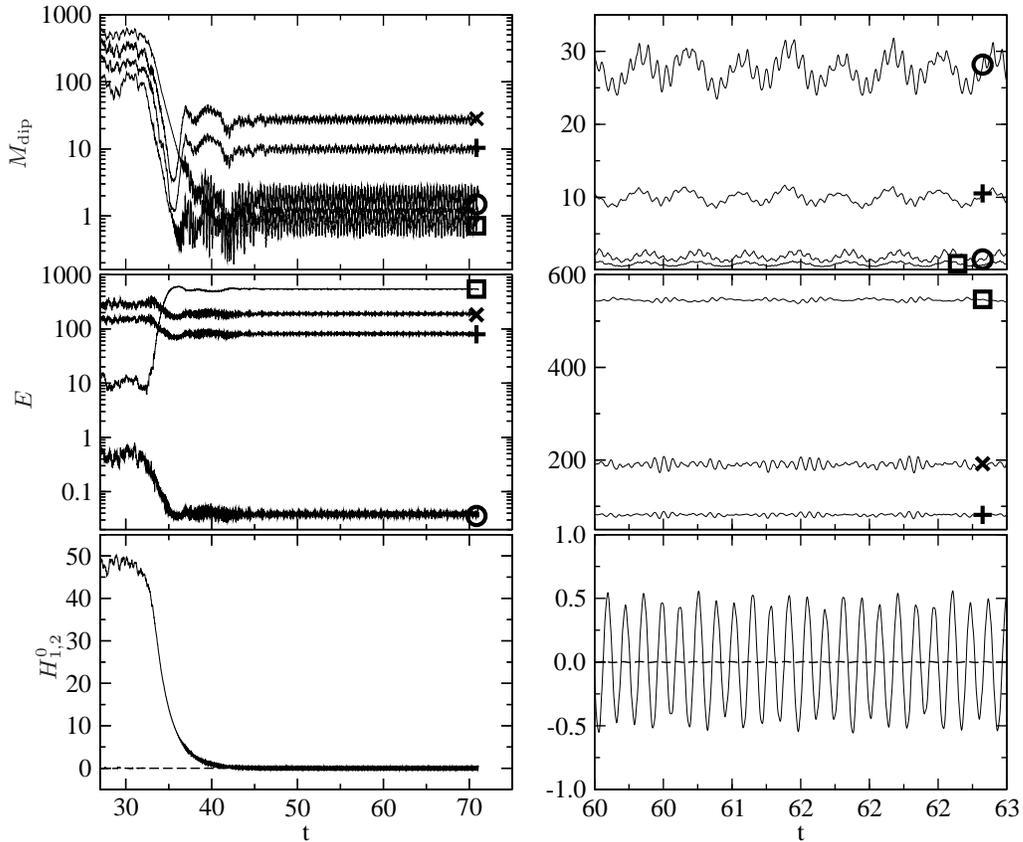}
\end{center}
\vspace{3mm}
\caption{Time series of a dynamo solution in the case
  $\P=1$, $\R=10^5$, $\tau=2000$, $\Pm=4.5$, $\eta=0.65$ and mixed
  velocity boundary conditions.
The first row shows dipolar magnetic energy densities. The second row
shows kinetic energy densities. The last row shows values of the
mean dipolar, $H_1^0$, and the mean quadrupolar, $H_2^0$, components in the
spherical harmonic expansion of the magnetic field. The components
$\overline{X}_p$ 
$\overline{X}_t$, $\widetilde{X}_p$, and $\widetilde{X}_t$ are indicated
by circles, squares, plus signs and crosses attached to the respective
curves,  and $X$ stands for either $M$ or $E$. The coefficient
$H_1^0$ is represented by a solid line and $H_2^0$ by a dashed
line. The time series in the right column represent enlarged sections of
the ones in the left column.
}
\label{fig03}
\end{figure}

\section{An attempt to model the solar magnetic cycle}

It appears from the results presented in the preceding section that
variations of the shell thickness alone are not sufficient to induce
oscillations in dipolar dynamo solutions. It is known that dipolar
oscillations may be excited in a variety of other situations
(Busse \& Simitev 2006). An oscillating dipolar solution can be found 
when a region in the parameter space is approached where the
quadrupolar or the higher-multipole components of the magnetic field
are not negligible. As these components are typically oscillatory, an 
oscillation in the dipolar parts is also excited. A region of
multipolar dynamos may be approached, for instance, by reducing the
value of the 
magnetic Prandtl number $\Pm$, or by increasing the value of the
Rayleigh number $\R$ or of the rotation parameter $\tau$. In all of these situations one finds
that the increase of the differential rotation plays a crucial
role in exciting the oscillations. The differential rotation may, of
course, be enhanced much more readily by imposing a stress-free
boundary condition for the velocity 
field. Thus, the suggestion of Goudard \& Dormy (2008) to replace the
no-slip condition on the outer spherical boundary by a stress-free one
is well-justified. As these authors remark, the choice of mixed
velocity boundary conditions may be argued also on physical grounds:
the no-slip condition at the inner boundary mimics the solar
tachocline (Tobias 2005) and the stress-free condition at the outer
boundary mimics the solar photosphere.

\begin{figure}
\vspace*{5mm}
\psfrag{Mdip}{$\mathrm{M}_\mathrm{dip}$}
\psfrag{E}{E}
\psfrag{t}{$t$}
\psfrag{HG}{$H_{1,2}^0$}
\begin{center}
%Data in
% /mnt2/dat/dyn/e065p1t2r100000m1p4.5mvbcB.period/l--plots--l/3mmm.txt
\epsfig{file=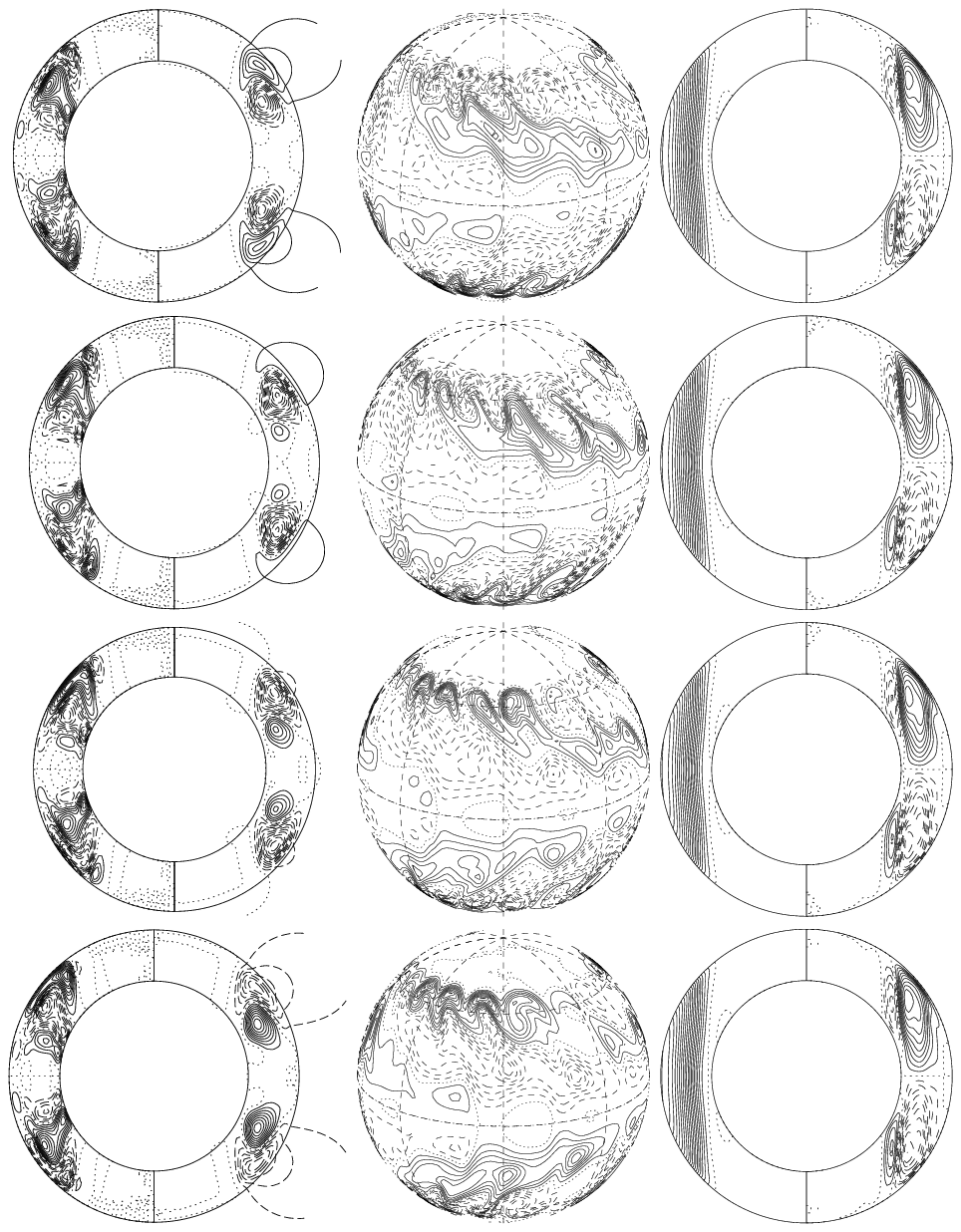,width=13cm,clip=}
\end{center}
\vspace{3mm}
\caption{Half a period of dipolar oscillations of the dynamo
  shown in Fig.~\ref{fig03}. 
  Each circle of the first column shows lines of constant
  $\overline{B_{\varphi}}$ in the left half and of $r \sin \theta
  \dd_\theta   \overline{h}=$ const.~in the right half. The second
  column shows meridional lines of constant $B_r$ at $r=r_o$. The
  third column shows meridional lines of constant
  $\overline{u}_\varphi$ in the left half and of $r \sin \theta
  \dd_\theta \overline{v}$ in the right half. 
%The last column shows
%  meridional lines of constant values of the azimuthally averaged
%  cross-helicity in the left half and helicity in the right
%  half. 
Plots follow from top to bottom with a time step $\Delta t= 0.021665$.
}
\label{fig04}
\end{figure}

In Fig.~\ref{fig03} a dynamo simulation with mixed velocity boundary
conditions is presented. After an interval of about 5 viscous
diffusion times an abrupt transition in the nature of the dynamo
solution occurs. The most notable signature of the transition is the
decrease of the magnetic energy density by more than an order of
magnitude. At the same time the kinetic energy increases twofold,
mostly due to the increase in differential rotation. This is not
surprising as the increase is promoted by two effects: first, by the
removal of the no-slip condition and second, by the decrease of
magnetic field which, now becomes too weak to effectively act as a
brake on differential rotation. This latter effect has been reported
in numerous other cases (Busse 2002, Simitev \& Busse 2005). 

More subtly, the transition is observed as a change in the relative
contributions of magnetic energy components. The 
mean poloidal dipolar magnetic energy $\overline{M}_p$ is no longer
the dominant component and is overcome by the fluctuating parts of the
magnetic energy. In this respect the transition is similar to the transition
between Mean Dipolar (MD) and Fluctuating Dipolar (FD) dynamos
recently reported by Simitev \& Busse (2009). 
\begin{figure}
\vspace*{-1cm}
\begin{center}
%Data in 
%/home/staff1/rs/07_DRS/CurrentRuns/e065p1t2r100000m1p4.5mvbcB.period/ButterfluDiagram.gnu
% %%BoundingBox: 100 150 500 720
\epsfig{file=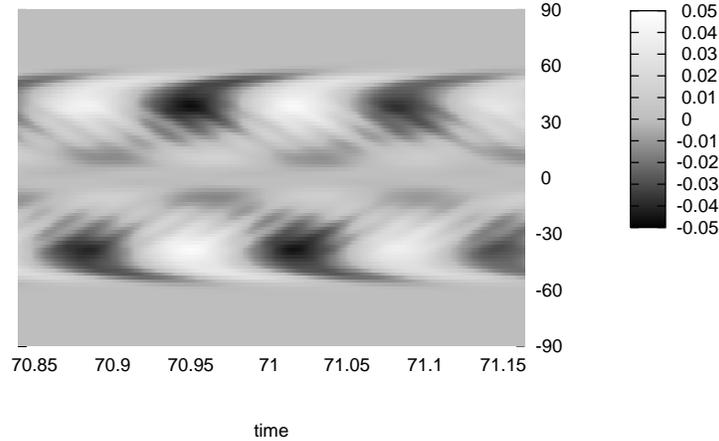,width=7cm,angle=-90,clip=}
\end{center}
\caption{A butterfly diagram for the case shown in
  Figs.~\ref{fig03} and \ref{fig04}, i.e.~the value of $r
  \partial_\theta \overline{g}$ at $r=r_o-0.05$ as a function of time
  and latitude.}  
\label{fig05}
\end{figure}

The temporal behavior of the dynamo changes dramatically after
the transition as well. In particular, it exhibits nearly
periodic dipolar oscillations. These are evident in the right-hand
side of Fig.~\ref{fig03} and even better in the oscillations of
the spherical harmonic coefficient $H_1^0$ describing the contribution
of the axial dipole in the solution. It is also remarkable that the
quadrupolar coefficient $H_2^0$ remains much weaker than in corresponding
oscillations in thick shell dynamos. A sequence of snapshots
equidistant in time is shown in Fig.~\ref{fig04} in order to visualize
half a period of oscillation. Except for the sign the plots shown in the last row closely
resemble the ones in the first row confirming thus the near perfect
periodicity of the oscillation. A polarity reversal of the magnetic field occurs
during the cycle and is evident both in Fig.~\ref{fig04} and in the
change of sign of $H_1^0$ in Fig.~\ref{fig03}. This cyclic
behavior is sustained for over 25 viscous diffusion times although it
appears from the right-hand side of Fig.~\ref{fig03} that a weak
modulation with a second frequency is also present.   Finally, we
remark that the oscillations appear to be well-described by the
Parker wave linear analysis presented in Busse \& Simitev (2006), in
that the period of oscillation $T=0.12996$ is well approximated by the
expression
$$
T \approx \left(P_m\frac{\pi}{3} \langle
\check{\vec v} \cdot \nabla \times \check{\vec v}
\rangle\sqrt{2\overline{E_t}}\right)^{-1/2}
$$
which in this case yields 0.11036.

In order to stress the similarity with the solar magnetic activity we
present in Fig.~\ref{fig05} a butterfly diagram of the simulation,
where the cyclic behavior and the propagation of magnetic features
from high latitudes toward the equator is also visible.

\section{Discussion and outlook}

We have confirmed the finding of Goudard \& Dormy (2008) of a sharp
transition from a dynamo dominated by a non-oscillatory dipolar
magnetic field to a nearly perfect oscillatory dynamo with a weaker
dipole. We have related this transition to corresponding transitions
from dynamos with a strong nearly steady dipole to oscillatory dynamos
in thick shells. The introduction of a stress-free boundary condition
at the outer spherical surface rather than the reduction of the shell
thickness appears to be essential for the transition. It is of
interest to note that the oscillations of thin  shell dynamos are much
more purely dipolar than those found in thick shells. 

It is tempting to compare the oscillatory dipolar dynamo of section
3 to the magnetic activity of the Sun. A notable difference with the
Sun is the profile of differential rotation of the solution which has
the form of a nearly constant angular velocity on
cylinders. Although this profile, when mapped to the surface $r=r_o$,
reproduces qualitatively the rotation of the solar surface,
the differences in radial direction are significant. 

This is not surprising in view of the strong change of density
throughout the solar convection zone caused by compressibility which
has not been taken into account in our simple model. In addition the
scales of turbulence are far from being resolved. This emphasizes the
need to explore the parameter space of the problem in further detail. 

\vspace*{0.2in}
We gratefully acknowledge support from CTR and NASA which made our
visit to Stanford possible. The numerical calculations reported in
this paper were carried out using the computer resources of the School of
Mathematics and Statistics of the University of Glasgow, and the UK
MHD supercomputer at the University of St.~Andrews, UK.

%\section*{References}
%
%uses the bibliography citations available with standard
%LaTeX, and formatting for a book and for a journal article are shown below.
%This is the raw code and the formatted output follows.  A reference
%can be cited as follows; \cite{Jang96}. 

%\bibliographystyle{jfm.bst}
%\bibliography{myrefs}

\begin{thebibliography}{15}
\expandafter\ifx\csname natexlab\endcsname\relax\def\natexlab#1{#1}\fi
 




%Schwabe, H.: 1844, Astron. Nachr. 21 (495), 233 Sonnen-Beobachtungen
%im Jahre 1843

\bibitem[Busse, 1970]{Busse1970}
{\sc Busse, F.H.} 1970 Differential rotation in stellar convection zones
{\em Astrophys. J.} {\bf 159} 629-639.

\bibitem[Busse, 2002]{Busse2002} {\sc Busse, F.H} 2002
Convective flows in rapidly rotating spheres and their dynamo action
{\em Phys Fluids} {\bf 14 (4)} 1301.

\bibitem[Busse \& Simitev, 2006]{Simitev2006} 
{\sc Busse, F.H. \& Simitev, R.} 2006 Parameter dependences of
convection driven dynamos in rotating spherical fluid shells
{\em Geophys. Astrophys. Fluid Dyn.} {\bf 100(4-5)} 341.

\bibitem[Brandenburg \& Subramanian, 2005]{Brandenburg}
{\sc Brandenburg, A. \& Subramanian, K.} 
Astrophysical magnetic fields and nonlinear dynamo theory.
{\em Physics Reports}, {\bf 417}, 1.


\bibitem[Browning, 2007]{Browning}
{\sc Browning, M., Brun, A.S., Miesch, M. \& Toomre, J.} 2007 Dynamo action in
simulations of penetrative solar convection with an imposed
tachocline {\em Astronom. Notes}  {\bf 328} 1002.

\bibitem[Brun \& Toomre, 2002]{Brun}
{\sc Brun, A. S. \& Toomre, J.} 2002 Turbulent convection under the
  influence of rotation: sustaining a strong differential rotation
{\em  Astrophys. J.} {\bf 570} 865.


\bibitem[Christensen et al., 2001]{Christensen}
{\sc Christensen, U.R., Aubert, J., Cardin P., \textit{et al.}} 2001 A
numerical dynamo benchmark {\em Phys. Earth Planet. Inter.} {\bf 128} 25.


\bibitem[Goudard \& Dormy, 2008]{Dormy}
{\sc Goudard, L. \& Dormy, E.} 2008 Relations between the dynamo
region geometry and the magnetic behavior
of stars and planets {\em EPL} {\bf 85} 59001.


\bibitem[R\"udiger \& Hollerbach, 2004]{Rudiger}
{\sc R\"udiger, G. \& Hollerbach, R.} 2004
Magnetic Universe: Geophysical And Astrophysical Dynamo Theory, 
{\em Wiley}

\bibitem[Schou et al., 1998]{Schou}
{\sc Schou, J., Antia, H., Basu, S., \textit{et al.}} 1998 Helioseismic studies
of differential rotation in the solar envelope by the solar
oscillations investigation using the Michelson Doppler Imager
{\em Astrophys. J.} {\bf 505} 390-417.



\bibitem[Simitev \& Busse, 2005]{Simitev2005}
{\sc Simitev, R. \& Busse, F.H.}  2005 Prandtl number dependence of
  convection driven dynamos in rotating spherical fluid shells
  {\em J. Fluid Mech.} {\bf 532} 365.

\bibitem[Simitev \& Busse, 2009]{Simitev2009}
{\sc Simitev, R. \& Busse, F.H.}  2009  Bistability and hysteresis of
dipolar dynamos generated by turbulent convection in rotating
spherical shells {\em  EPL} {\bf 85} 19001.


\bibitem[Tilgner, 1999]{Tilgner}
{\sc Tilgner, A.} 1999 Spectral methods for the simulation of
incompressible flows in spherical shells
{\em Int. J. Num. Meth. Fluids} {\bf 30 (6)} 713.

\bibitem[Tobias, 2005]{Tobias}
{\sc Tobias, S.} 2005 The solar tachocline: Formation, stability and its
role in the solar dynamo, pp. 193-233 in {\em Fluid Dynamics and Dynamos in
Astrophysics and Geophysics}, A.M.Soward, C.A.Jones, D.W.Hughes,
N.O.Weiss (eds.), CRC Press.


\end{thebibliography}

\end{document}